\documentclass[prl,twocolumn,reprint,superscriptaddress]{revtex4}
\usepackage{amsmath}		%使用数学模式
\usepackage{amssymb}		%支持更多数学符号
\usepackage{mathtools}
\usepackage{mathrsfs}
\usepackage{graphicx}		%加入插图
\usepackage{subfigure}		%子插图
\usepackage{physics}		%快速输入物理学中常用符号
\usepackage{tensor}			%输入张量
\usepackage{verbatim}
\usepackage{hyperref}
\usepackage{xcolor}

\newcommand{\ii}{\mathrm i}

\allowdisplaybreaks[2]

\begin{document}

\title{Non-quasiconvex dispersion of composite fermions and the
  fermionic Haffnian state in the first-excited Landau level}

\author{Hao Jin}

\affiliation{International Center for Quantum Materials, Peking
  University, Beijing 100871, China}

\author{Junren Shi}
\email{junrenshi@pku.edu.cn}

\affiliation{International Center for Quantum Materials, Peking
  University, Beijing 100871, China}

\affiliation{Collaborative Innovation Center of Quantum Matter,
  Beijing 100871, China}

\begin{abstract}
  It has long been puzzling that fractional quantum Hall states
  in the first excited Landau level (1LL) often differ
  significantly from their counterparts in the lowest Landau
  level.  We show that the dispersion of composite fermions (CFs)
  is a deterministic factor driving the distinction.  We find
  that CFs with two quantized vortices in the 1LL have a
  non-quasiconvex dispersion.  Consequently, in the filling
  fraction $7/3$, CFs occupy the second $\Lambda$-level instead
  of the first.  The corresponding ground state wave function,
  based on the CF wave function ansatz, is identified to be the
  fermionic Haffnian wave function rather than the Laughlin wave
  function.  The conclusion is supported by numerical evidence
  from exact diagonalizations in both disk and spherical
  geometries.  Furthermore, we show that the dispersion becomes
  quasiconvex in wide quantum wells or for CFs with four
  quantized vortices in the filling fraction $11/5$, coinciding
  with observations that the distinction between the Landau
  levels disappears under these circumstances.
\end{abstract}

\maketitle

\paragraph{Introduction.---} A two-dimensional electron gas
subjected to a strong perpendicular magnetic field exhibits the
fractional quantum Hall effect (FQHE), characterized by
fractionally quantized Hall plateaus in specific filling
fractions of Landau levels~\cite{PhysRevLett.48.1559,RevModPhys.71.875}.  The effect is also observed in
topological flat bands~\cite{neupert_fractional_2015} in recent experiments~\cite{cai2023signatures,park2023,xu2023,zeng2023}.
Theoretical understanding of the FQHE is challenging because the
kinetic energies of electrons in a Landau level are completely
quenched, leaving interaction to dominate.  Consequently,
constructing plausible ground state wave functions for the FQHE
has long relied on intuition or educated guesses, from the
celebrated Laughlin wave function~\cite{PhysRevLett.50.1395} to those hypothesized by
more elaborate approaches such as the hierarchy theory~\cite{PhysRevLett.51.605,PhysRevLett.52.1583}, the
conformal field theory~\cite{CRISTOFANO199188,doi:10.1142/S0217732391000336}, and the composite fermion (CF)
theory~\cite{PhysRevLett.63.199,jain2007composite}.  Among these, Jain's CF theory is the most
successful.  The theory introduces fictitious particles called
CFs, each consisting of an electron and an even number of
quantized vortices.  Wave functions prescribed by the CF theory
yield nearly perfect overlaps with those obtained from exact
diagonalizations (ED) in the lowest Landau level (LLL)~\cite{PhysRevB.45.1223,wu_mixed-spin_1993,jain1997}.
Furthermore, it has been shown that a deductive approach for
determining CF wave functions and corresponding physical wave
functions can be established~\cite{shi2023quantum}.

While the FQHE in the LLL can be well described by the CF theory,
it has long been puzzling that the FQHE in the first excited
Landau level (1LL) often exhibits distinct features.  Most
notably, a Hall plateau is developed in the even denominator
fraction $\nu=5/2$~\cite{PhysRevLett.59.1776}, which has been a focus in the pursuit of
topological quantum computing~\cite{RevModPhys.80.1083}.  On the other hand, ordinary
fractions with odd denominators in the 1LL, such as $\nu=7/3$ and
$\nu=12/5$, differ significantly from their counterparts in the
LLL~\cite{d1988fractional,PhysRevB.63.125312,WOJS200263,PhysRevB.72.125315}.  The observation casts doubt on the applicability of the
CF theory in the 1LL, motivating alternative pictures such as the
parafermion theory~\cite{PhysRevB.59.8084} and the parton theory~\cite{PhysRevB.40.8079,PhysRevResearch.2.013349}.

A representative case is the filling fraction $\nu=7/3$ with an
effective filling fraction $\tilde\nu = 1/3$ in the 1LL~\cite{PhysRevLett.59.1776,d1988fractional}.
This state is expected to be an analog of the $1/3$ state in the
LLL and described by the $1/3$ Laughlin wave function.  However,
it is found that the overlap between the $1/3$ and $7/3$ states
is low~\cite{d1988fractional,PhysRevB.72.125315}, and their entanglement and quasi-hole excitation
spectra are distinct~\cite{PhysRevB.91.045115,PhysRevLett.110.186801,PhysRevB.89.115124}.  More puzzlingly, ED shows that the
$11/5$ state can nevertheless be well described by the $1/5$
Laughlin wave function~\cite{d1988fractional,PhysRevB.72.125315}.  Several theories had been put
forward to explain the peculiarity of the $7/3$ state.  T\"{o}ke \emph{et al.}
suggest that there exists substantial $\Lambda$-level mixing
induced by residue interaction between CFs~\cite{PhysRevB.72.125315}.  Balram \emph{et al.}
propose that the $7/3$ state hosts
$\mathbb{Z}_n$-superconductivity of partons~\cite{PhysRevResearch.2.013349}.  Various trial
wave functions for this fraction have been numerically tested in
Refs.~\onlinecite{PhysRevB.63.125312,WOJS200263,PhysRevB.96.125148,PhysRevB.97.245125}.

In this Letter, we demonstrate that the dispersion of CFs is a key
for understanding the peculiarities of the 1LL.  Based on the
deductive approach developed in Ref.~\onlinecite{shi2023quantum},
we show that CFs with two quantized vortices (denoted as CF$^2$)
in the 1LL, unlike CFs in the LLL, have a non-quasiconvex
dispersion.  Consequently, in $\nu=7/3$, CFs occupy the second
$\Lambda$-level~\cite{jain2007composite} instead of the first.  The corresponding
ground state is identified to be the fermionic Haffnian state,
also known as the $d$-wave paired FQH state of spinless
electrons~\cite{WEN1994455,yang2021}, rather than the Laughlin state.  Using ED, we
demonstrate that the unique features of the Haffnian state do
manifest in exact ground state wave functions.  Furthermore, we
find that the dispersion becomes quasiconvex in wide quantum
wells or for CF$^4$ in the $11/5$ state, coinciding with
observations that the distinction between the 1LL and LLL
disappears under these circumstances.  These findings,
cumulatively, support that the non-quasiconvex CF dispersion is a
deterministic factor behind the peculiarities of the FQHE in the
1LL.

\paragraph{CF dispersion.---} A CF consists of an electron and a
vortex carrying an even number of quantized vortices.  The
Coulomb attraction between the electron and the charge void
induced by the vortex gives rise to the binding energy of the
CF~\cite{shi2023quantum}.  Read shows that the momentum of a CF can be defined as
being proportional to the spatial separation between the electron
and vortex~\cite{read1994theory}.  Consequently, the binding energy, as a function
of the spatial separation $r$, can be interpreted as the
dispersion of the CF~\cite{ji2021}.  To determine the binding energy, we
first calculate the electron-vortex correlation function
$h\qty(r)$, which describes the electron density profile in the
vicinity of a vortex.  The binding energy can then be determined
by $\epsilon_b\qty(r)= \qty(\rho_0/2)\int\dd[2]{\boldsymbol{r}'}v
\qty(\abs{\boldsymbol{r}-\boldsymbol{r}'})h\qty(r')$, where
$v\qty(r)$ is the interaction between electrons and $\rho_0$ is
the average electron density~\cite{shi2023quantum}.

\begin{figure}[t]
	\centering
	\includegraphics[width=\linewidth]{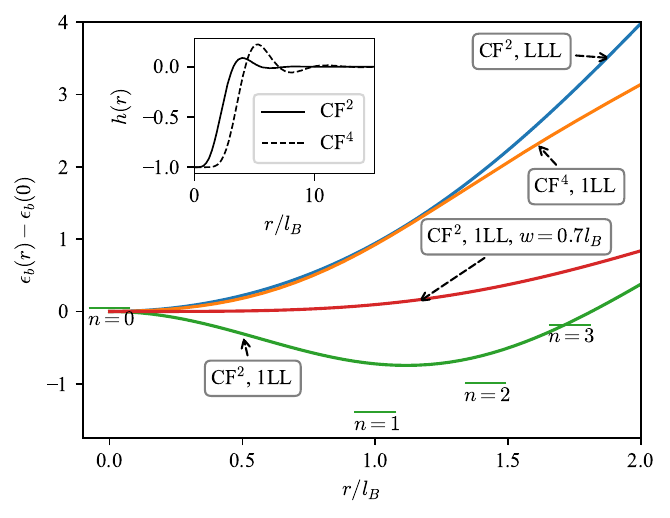}
  	\caption{Binding energy $\epsilon_{b}\qty(r)$ for CF$^2$ and
          CF$^4$ in the LLL and 1LL.  The binding energy of
          CF$^2$ in the 1LL in a quantum well with a finite width
          $w=0.7 l_B$ is also shown. The horizontal line segments
          indicate the energies of $\Lambda$-levels for $\nu=7/3$.  The
          energies are in units of $\tilde{\nu} e^2/16\pi^{2} \epsilon l_B$.  Inset:
          electron-vortex correlation function $h\qty(r)$ for
          CF$^2$ and CF$^4$.\label{fig_dispersion_eig}}
  
\end{figure}

We can determine $h\qty(r)$ for CF$^2$ by assuming the ground
state to be the $1/3$ Laughlin state.  The many-body wave
function in the presence of a vortex at the origin is given by
$\Psi_0^\mathrm{v}\qty(\qty{z_i})=\prod_{i}z_i^2\prod_{i<j}\qty(z_i-z_j)^3$,
where $\qty{z_i=x_i+y_i}$ is the set of complex electron
coordinates~\cite{shi2023quantum}.  $h\qty(r)$ is obtained by computing the
electron density distribution of $\Psi_0^{\mathrm v}$ normalized
by $\rho_0$.  The result is shown in the inset of Fig. \ref{fig_dispersion_eig}.

Different interactions $v(r)$ give rise to different binding
energies of CF$^{2}$ for the LLL and the 1LL.  For the LLL,
$v(r)$ is the Coulomb interaction
$v_\mathrm{c}\qty(r)=e^2/4\pi\varepsilon r$.  For the 1LL, we map
the problem of interacting electrons to the mathematically
equivalent problem of interacting electrons in the LLL with the
effective interaction~\cite{jain2007composite}:
\begin{equation}
\tilde{v}\qty(r)=\qty(1+\frac{l_B^2\nabla^2}{2})^2
v_\mathrm{c}\qty(r),\label{eq_tildeV}
\end{equation}
where $l_B=\sqrt{\hbar/eB}$ is the magnetic length with $B$ being
the strength of the magnetic field.  The binding energy of
CF$^2$ in the LLL and 1LL can then be determined using $h\qty(r)$ and
the respective interactions.  The results are shown in Fig.~\ref{fig_dispersion_eig}.
We observe that the binding energy of CF$^{2}$ in the 1LL, unlike
that in the LLL, is a non-quasiconvex function of $r$.

\begin{comment} Similarly, we can determine the binding energy of CF$^4$, which is relevant to the filling fraction $\nu=11/5$. $h\qty(r)$ is obtained from the wave function $\Psi_0^\mathrm{v}\qty(\qty{z_i})=\prod_{i}z_i^4\prod_{i<j}\qty(z_i-z_j)^5$.  We find that the resulting dispersion is a quasiconvex function of $r$, similar to that of CF$^2$ in the LLL.
\end{comment}

\paragraph{$\Lambda$ levels and wave functions.---} With the
binding energy of a CF, we can establish its effective
Hamiltonian.  In the dipole picture, the electron and vortex in a
CF are confined in two separate LLLs induced by the physical
magnetic field and the emergent Chern-Simons magnetic field,
respectively~\cite{shi2023quantum}.  As a result, the CF is described by a
bi-variate wave function $\psi\qty(z,\bar\eta)$, which is
(anti-)holomorphic in the complex electron (vortex) coordinate
$z$ ($\eta\equiv\eta_x+\ii\eta_y$).  After projecting to the
LLLs, $\bar z$ and $\eta$ become the operators
$\hat{\bar z}=2l_B^2\partial_z$ and
$\hat\eta=2l_b^2\partial_{\bar\eta}$, respectively, where
$l_b=l_B/\sqrt{2\tilde\nu}$ is the magnetic length of
Chern-Simons magnetic field for
CF$^{2}$~\cite{jain2007composite,shi2023quantum}.  We can then define the ladder operators of
$\Lambda$-levels as $\hat a=\qty(z-\hat\eta)/\sqrt\gamma l_B$ and
$\hat a^\dagger=\qty(\hat{\bar z}-\bar\eta)/\sqrt\gamma l_B$,
with $\gamma\equiv\abs{1/\tilde\nu-2}=1/q$ for the filling
fraction $\tilde\nu=q/(2q+1)$.  The effective Hamiltonian
operator of the CF is given by:
\begin{equation} \hat H=:\epsilon_b\qty(\hat r):,
\end{equation}
where we express the binding energy as a function of
$r^2\equiv\abs{z-\eta}^2$, which is then mapped to the operator
$\hat r^2\equiv\gamma l_B^2\hat a^\dagger\hat a$.  The colons
indicate the normal ordering of the ladder operators, which
places $\hat a^\dagger$s to the left of $\hat a$'s.  The specific
ordering of the ladder operators is a result of the particular
way that the binding energy is defined~\cite{shi2023quantum}.

The eigen-energies and eigen-states of $\hat H$ can be
determined.  It is easy to see that the number operator
$\hat n\equiv\hat a^\dagger\hat a$ commutes with the Hamiltonian:
$[\hat H,\hat n]=0$.  We can thus define the index of
$\Lambda$-levels as the eigenvalue of $\hat{n}$.  Similar to
ordinary Landau levels, wave functions for the first ($n=0$)
$\Lambda$-level is annihilated by the lowering operator $\hat a$.
We thus have
$\psi_{0}\qty(z,\bar\eta)=f(z)\exp(z\bar{\eta}/2l_{b}^{2})$,
where $f(z)$ is a holomorphic function in $z$.  Wave functions
for $n>0$ can be obtained by successively applying the raising
operator to $\psi_{0}\qty(z,\bar\eta)$:
$\psi_{n}\qty(z,\bar\eta)=(\hat a^{\dagger
  n}/\sqrt{n!})\psi_{0}\qty(z,\bar\eta)$.  The energies of
$\Lambda$-levels as a function of $n$ can be determined
straightforwardly~\cite{10.21468/SciPostPhys.10.1.007}.  The result for
$\nu=7/3$ is shown in Fig.~\ref{fig_dispersion_eig}.  Notably, the lowest
$\Lambda$-level has the index $n=1$ rather than $n=0$.

The physical wave function of the $7/3$ state can be determined
by applying the wave function ansatz of the CF theory, which maps
a many-body wave function of CFs to a physical wave function of
electrons.  From the energy spectrum of $\Lambda$-levels, we
expect that CFs in the $7/3$ state fully occupy the second
$\Lambda$-level.  The corresponding wave function of CFs can be
obtained by applying raising operators to the wave function
$\Psi_{\text{CF}}^0\qty(\qty{z_i,\bar\eta_i})$ of a fully
occupied first $\Lambda$-level:
$\Psi_{\text{CF}}\qty(\qty{z_i,\bar\eta_i})\propto\prod_{i}
\qty(2l_B^2\partial_{z_i}-\bar\eta_i)\Psi^0_{\text{CF}}\qty(\qty{z_i,\bar\eta_i})$,
with
$\Psi_{\text{CF}}^0\qty(\qty{z_i,\bar\eta_i})\equiv
\prod_{i<j}\qty(z_i-z_j)\exp(\sum_i {z_i\bar\eta_i}/{2l_b^2})$.
The electron wave function can then be obtained by overlapping
$\Psi_{\text{CF}}$ with the $1/2$ Laughlin state of vortices~\cite{shi2023quantum}.
We obtain~\cite{supp}:
\begin{equation} \Psi(\qty{z_i}) \propto\lim\limits_{\eta\to
z}\prod_{i}(2\partial_{z_i}-\partial_{\eta_i})
\prod_{i<j}\qty(z_i-z_j)\qty(\eta_i-\eta_j)^2.\label{eq_wfc_Lambda2}
\end{equation}

We can obtain an explicit form of the wave function~\cite{supp}.
For an even number of electrons, it is the fermionic Haffnian
wave function~\cite{greiter1992paired,WEN1994455,green2002stronglycorrelatedstateslow}:
\begin{eqnarray}
\Psi\qty(\qty{z_i})\propto\text{Hf}\qty[\frac{1}{\qty(z_i-z_j)^2}]
\prod_{i<j}\qty(z_i-z_j)^3,\label{eq_Hafnian}
\end{eqnarray}
where $\text{Hf}$ denotes the Haffnian of a matrix with
off-diagonal elements $1/\qty(z_i-z_j)^2$.  The specific Haffnian
can also be written as the determinant
$\det[1/\qty(z_i-z_j)]$~\cite{greiter1992paired}.  For an odd number of electrons, on
the other hand, the wave function Eq.~(\ref{eq_wfc_Lambda2}) predicted by the CF
theory is identically zero~\cite{wu1995}.  The Haffnian state was first
proposed as a $d$-wave pairing state analogous to the $p$-wave
Pfaffian state~\cite{WEN1994455}.  As far as we know, this is the first time
that the state is related to the CF theory, with the underlying
CF state identified.

Similarly, we can determine the $\Lambda$-levels for the filling
fraction $\nu=12/5$ with the effective filling fraction
$\tilde\nu=2/5$ in the 1LL.  In this case, we find that the two
lowest $\Lambda$-levels have the indices $n=2,\,3$~\cite{supp}.
Consequently, CFs in the $12/5$ state occupy the third and fourth
$\Lambda$-levels rather than the first two as in the $2/5$ state.
The resulting electron state should also differ significantly
from the $2/5$ state in the LLL.
\begin{comment} In above calculation, we start with the assumption that $7/3$ is a Laughlin state, then we compute $h\qty(r)$ and the energies of $\Lambda$ levels.  It turns out that $n=1$ $\Lambda$ level is the lowest and $7/3$ is a fermionic Hafnian state, conflict with the assumption.  For consistency, we re-calculate these quantities based on the assumption that $7/3$ is fermionic Hafnian state.  $h\qty(r)$ is computed using the wave function $\Psi_0^\mathrm{v}\qty(\qty{z_i})=\det\qty(P_{ij})\prod_{i}z_i^2 \prod_{i<j}\qty(z_i-z_j)^3$, where $P_{ij}=1/\qty(z_i-z_j)$ for $i\neq j$ and $P_{ii}=-\sqrt{2/3}/z_i$.  The detail is discussed in supplemental material.  We find that $n=1$ $\Lambda$ state is still the lowest.  In this way we predict that $7/3$ is fermionic Hafnian state.
\end{comment}

\paragraph{Numerical verification.---} Using ED, we test our
conclusion for $\nu=7/3$ by examining whether features of the
Haffnian state manifest in exact ground state wave functions.
Compared to the Laughlin state, the Haffnian state has a smaller
$z$-component of the total angular momentum $L_{z}$ on a disk and
a different topological shift $S$ on a sphere.  Moreover, the
pairing nature of the state suggests stability only for even
numbers of electrons.  We therefore solve the ground state wave
functions of interacting electrons in both the disk and spherical
geometries.  The results corroborate our conclusion well.

\begin{figure}[t]
\centering \includegraphics[width=\linewidth]{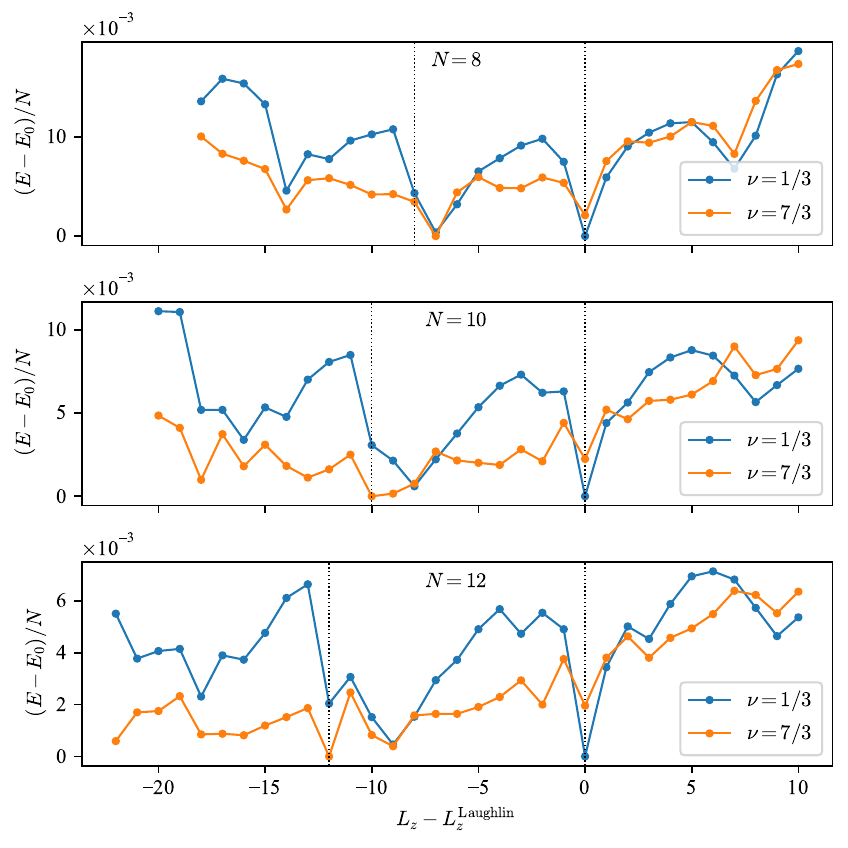}
\caption{Lowest ED energies per particle of $\nu=1/3$ and $\nu=7/3$
  in different $L_z$ sectors.  $E_0$ denotes the ground state
  energy.  The energies are in units of $e^2/4\pi\epsilon l_B$.
  $L_z^\text{Laughlin}=3N(N-1)/2$ denotes the $z$ component of
  the total angular momentum of the Laughlin state.  The vertical
  dashed lines indicate the values expected for the Haffnian
  state $L_z^\text{Haffnian}=L_z^\text{Laughlin}-N$.  }
  \label{fig_disk_energy}
\end{figure}

In the disk geometry, $L_z$ of the Haffnian state Eq.~\eqref{eq_Hafnian} is
smaller than that of the Laughlin state by $N$, where $N$ is the
number of electrons.  Figure~\ref{fig_disk_energy} shows the lowest eigen-energies
in different $L_{z}$ sectors for $\nu=1/3$ and $7/3$.  It is evident
that the exact ground states for $\nu=7/3$ do have the $L_z$
values expected for the Haffnian state for $N\geq10$.  For $N=8$,
we observe a deviation by one, likely due to finite size effects.
\begin{comment} We put $N$ electrons on a uniformly charged disk centered at origin, which carries charge $Ne$ and charge density $e\rho_0$.  The upper limit of $z$ component angular momentum of each electron is $3N-1+D$ where $D$ is a positive integer.  We emphasize that, different from Ref\cite{PhysRevB.72.125315}, the filling factor in our computation is specified through the charge density of background.  We perform ED computation for $\nu=1/3$ and $7/3$ with $D=5$.  The total energy consists of three parts: (1) the electron-electron interaction $V_{el-el}$, (2) the potential felt by electrons $\Phi$, (3) the self energy of charged background $V_{bg-bg}$.  We note that $\Phi$ is the Coulomb potential $\Phi^c\qty(r)$ in the LLL while $\Phi=\qty(1+l_B^2\nabla^2/2)\Phi^c\qty(r)$ in the 1LL.
\end{comment}

In the spherical geometry, the Haffnian state has a different
topological shift $S$ compared to the Laughlin state.  The
topological shift $S$ is defined by the relation
$2l\equiv\tilde{\nu}^{-1}N-S$, where $l$ is the angular-momentum
quantum number of the physical Landau level~\cite{PhysRevLett.69.953}.  For CF$^{2}$,
we have $l=l^{\ast}+(N-1)$, where $l^{\ast}$ is the
angular-momentum quantum number of the first
$\Lambda$-level~\cite{jain2007composite,PhysRevLett.51.605}.  Fully occupying a $\Lambda$-level with the
index $n$ requires $N=2\qty(l^\ast+n)+1$.  Combining these
relations, we have $S=3$ and $S=5$ for the Laughlin state ($n=0$)
and the Haffnian state ($n=1$), respectively.
\begin{comment} A magnetic monopole with strength $2q>0$ is put at the center of the sphere, which generates isotropic magnetic field and LLs.  The LLL on sphere has degeneracy $2q+1$ \cite{jain2007composite}.  The wave functions on disk are mapped to the sphere through stereographic projection \cite{DUNNE1992233}.
\end{comment}

While $S$ is a free parameter for ED calculations in the
spherical geometry, its probable value could be identified by
examining the degeneracies and stability (excitation gaps) of the
ground states with respect to different values of $S$ and $N$.
For the $7/3$ state, W\'{o}js \emph{et al.} investigated a few candidate
values of $S$ and concluded that $S=7$ is the most probable,
based on the reasoning that for the particular shift, a gaped
non-degenerate ground state with a total angular momentum $L=0$
can always be found for all calculated values of $N$~\cite{WOJS200263,PhysRevB.63.125312}.  $S=5$
was ruled out because the ground states become degenerate ($L>0$)
for odd numbers of electrons.

\begin{figure}[t]
\centering \includegraphics[width=\linewidth]{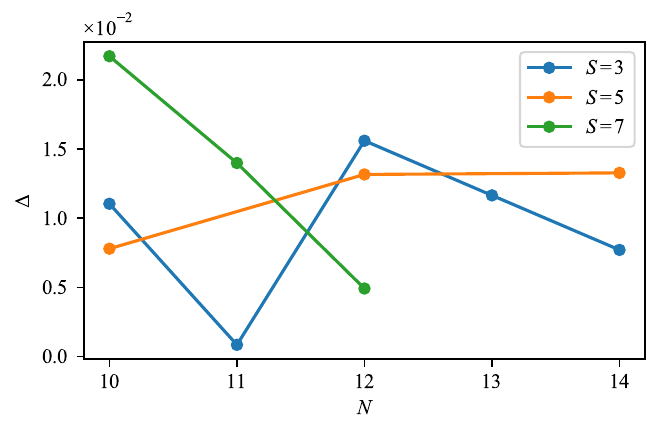}
\caption{Excitation gaps for non-degenerate ground states at
  $S=3,\,5,\,7$, in units of $e^2/4\pi\epsilon l_B$.}
  \label{fig_gap_S}
\end{figure}

To this end, we repeat the calculation and extend it for larger
values of $N$.  In Fig.~\ref{fig_gap_S}, we show the $N$ dependence of the
excitation gaps for non-degenerate ground states at
$S=3,\,5,\,7$.  Our calculation confirms W\'{o}js \emph{et al.}'s
observation that the excitation gap for $S=7$ diminishes rapidly
with increasing $N$ for $N\leq12$.  When $N$ is further
increased, we observe that the ground states become degenerate.
Conversely, although the ground states for $S=5$ are
non-degenerate only for even numbers of electrons, the magnitude
of the excitation gap shows a trend of converging to a constant
value for $N\geq12$, suggesting robustness of the non-degenerate
ground states for $S=5$.
\begin{comment} various wave functions with different values of $S$ have long been proposed~\cite{PhysRevB.72.125315,PhysRevLett.110.186801,PhysRevResearch.2.013349,PhysRevB.96.125148}.
\end{comment}
\begin{comment} (With $S=3$, Jain interprets the $7/3$ state as a state with strong CF interaction \cite{PhysRevB.72.125315,PhysRevLett.110.186801} and Balram constructs a parton wave function for $7/3$ state \cite{PhysRevResearch.2.013349}.  Other trial wave functions with $S=-3,5$ are studied in Ref. \cite{PhysRevB.96.125148}. )
\end{comment}

\begin{figure}[htb]
\centering \includegraphics[width=\linewidth]{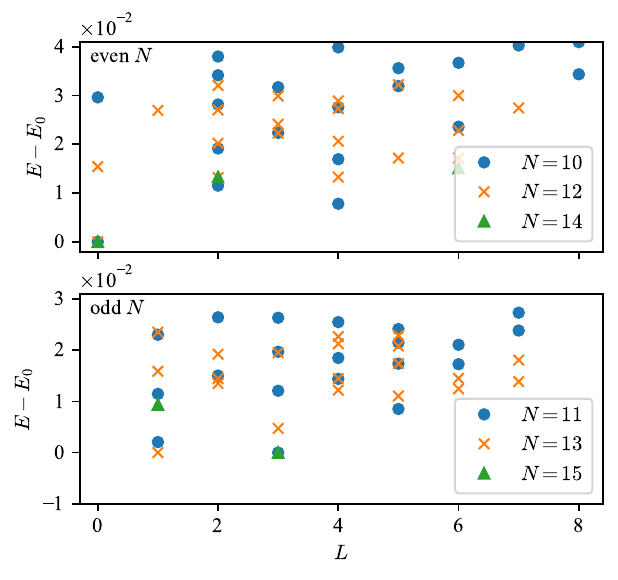}
\caption{ED spectrum for even and odd numbers of electrons.  $E_0$
  denotes the ground state energy.  The energies are in units of
  $e^2/4\pi\epsilon l_B$.  For $N=14$ ($N=15$), only three (two) lowest
  levels are shown.}
  \label{fig_E_N}
\end{figure}

In retrospect, we believe that the odd-even alternation observed
in the ED results for $S=5$ should be interpreted as a
manifestation of the pairing nature of the Haffnian state, rather
than a reason to dismiss it.  Remarkably, the CF theory, which
projects the ground state of CFs for odd $N$ to zero
identically~\cite{wu1995}, predicts the alternation.  With the CF ground
state annihilated, the low-lying excited states of CFs are
expected to generate both the ground state and low-lying excited
states of electrons for odd $N$ with $S=5$.  In Fig.~\ref{fig_E_N}, we
present the low-lying electron energy spectrum for $S=5$.  We do
observe that the spectrum for odd $N$ bears resemblance to the
excited portion of the spectrum for even $N$.

\paragraph{Non-quasiconvex to quasiconvex transition.---} To
further corroborate that the dispersion of CFs is a deterministic
underlying factor, we investigate two scenarios under which the
CF dispersion in the 1LL becomes quasiconvex.  We anticipate that
the distinction between the 1LL and LLL should disappear when the
dispersion becomes quasi-convex.  This is indeed observed.

The first scenario involves a different $h\qty(r)$.  We consider
CF$^4$, which carries four quantized vortices and underlies the
filling fraction $11/5$.  In this case, the binding energy can be
determined similarly to CF$^{2}$, albeit using the wave function
$\Psi_0^\mathrm{v}\qty(\qty{z_i})=\prod_{i}z_i^4\prod_{i<j}\qty(z_i-z_j)^5$.
We find that the resulting binding energy is a quasiconvex
function of $r$, similar to that of CF$^2$ in the LLL, as shown
in Fig.~\ref{fig_dispersion_eig}.  This is consistent with the fact that the $11/5$
state can be well described by the Laughlin wave function~\cite{d1988fractional,PhysRevB.72.125315}.

The second scenario involves the change of the electron
interaction $v(r)$.  We adopt the modified Coulomb interaction
investigated in Ref.~\onlinecite{papic_interaction-tuned_2009}:
$v_{\mathrm{c}}\qty(r)=e^2/4\pi\epsilon\sqrt{r^2+w^2}$, where $w$
is a parameter for characterizing the width of the quantum well
confining the two-dimensional system.  We find that for CF$^{2}$
in the 1LL, the dispersion becomes quasiconvex for
$w\gtrsim 0.7l_B$, as shown in Fig.~\ref{fig_dispersion_eig}.  This is consistent with
the ED results of Ref.~\onlinecite{papic_interaction-tuned_2009},
which show that the overlap between the exact ground state wave
function of the $7/3$ state and the $1/3$ Laughlin wave function
approaches unity for $w \gtrsim l_{B}$.

\paragraph{Summary and discussion.---} In summary, we show that
CF dispersion is a deterministic factor behind the peculiarities
of the FQHE in the 1LL.  The deductive approach enables us to not
only explain the distinction between the $7/3$ or $12/5$ states
and their counterparts in the LLL, but also correctly predict the
evolution of electron states under varying conditions.

Remarkably, the Haffnian state, considered analogous to the
Pfaffian pairing state, can be linked to a non-interacting CF
state.  The non-interacting nature also suggests that
quasi-electron or hole excitations of such a Haffnian ground
state should resemble those of the Laughlin state~\cite{toke2009}, rather
than those expected from conformal field theory and model
Hamiltonian considerations~\cite{hermanns2011,yang2021}.  This interpretation seems to be
consistent with the results of tunneling measurements~\cite{dolev2010,baer2014}.

We need to point out that the single-particle mean-field approach
employed in this study may not be adequate in fully capturing
quantitative details of a state with strong pairing correlations,
as in the case of the Haffnian state.  Numerically, the overlap
between the Haffnian wave function and the exact ground state
wave function for $\nu=7/3$ is found to be moderate~\cite{PhysRevB.97.245125}.  The
pairing correlations also complicate the self-consistent
determinations of the electron-vortex correlation function and
the CF dispersion~\cite{supp}.  Nevertheless, as pointed out in
Ref.~\onlinecite{jain2007composite}, it is not uncommon for a
qualitatively correct wave function to yield a low overlap with
the exact wave function.  It is reasonable to expect that
considering the residual interaction between CFs could improve
the quantitative agreement~\cite{PhysRevB.72.125315,PhysRevLett.127.046402}.

\begin{acknowledgments}
  We acknowledge B. Yang and X. Lin for useful discussions.  This
  work is supported by the National Key R\&D Program of China
  under Grand No.~2021YFA1401900 and the National Science
  Foundation of China under Grant No.~12174005.
\end{acknowledgments}

\begin{comment} one can obtain linear relation between the excitation gap and filling factor, which fits well with experiment data \cite{jain2007composite,PhysRevLett.70.2944,PhysRevLett.73.3270}. However this fails for the 1LL\cite{PhysRevLett.105.246808}.  The non-quasiconvex CF dispersion will lead to non-uniform gaps between $\Lambda$ levels, as shown in Fig\ref{fig_dispersion_eig}(b).  For example, the $n=0,3,4$ $\Lambda$ levels are close to each other, making corresponding FQHE gap small or unobservable.  This may explain the absent of $\nu=2+3/7,2+4/9$ FQHE in experiment\cite{PhysRevLett.105.246808,PhysRevB.77.075307}.  Another important issue is the origin of $\nu=5/2$ FQHE\cite{PhysRevLett.94.166802,RevModPhys.80.1083}.  It is widely believed that this is caused by pairing between CFs, while the details of pairing is not clear and various candidate wave functions are proposed \cite{MOORE1991362,PhysRevLett.99.236806,PhysRevLett.99.236807,PhysRevX.5.031027,PhysRevB.98.115107}.  These wave functions claim different pairing channel, $p+\ii p,p-\ii p$ or even $s$ wave.  For non-quasiconvex CF dispersion like Fig\ref{fig_dispersion_eig}(a), there will be two CF fermi surfaces.  The pairing is expected to be amplified by the two fermi surface structure and the scattering between two fermi surfaces would be vital to determine the pairing channel.
\end{comment}

\bibliographystyle{apsrev4-2}

\bibliography{ref}

\end{document}

\title{Supplemental Material for Non-quasiconvex dispersion of composite
  fermions and the fermionic Hafnian state in the first-excited Landau level}
\author{Hao Jin}

\affiliation{International Center for Quantum Materials, Peking University,
  Beijing 100871, China}

\author{Junren Shi}

\affiliation{International Center for Quantum Materials, Peking University,
  Beijing 100871, China}

\affiliation{Collaborative Innovation Center of Quantum Matter, Beijing 100871,
  China}

\maketitle

\section{Notations}

\subsection{Bergman space}
A wave function in the LLL can generally be written as $\psi(z) \exp
(-|z|^{2}/4l_{B}^{2})$, where $\psi(z)$ is a holomorphic function of the
particle coordinate $z\equiv x + \mathrm{i} y$, and $l_{B}\equiv\sqrt{\hbar/eB}$
denotes the magnetic length.

Since the Gaussian factor is common for all states in the LLL, it is convenient
to represent the states using only the holomorphic part of the wave function.
In this work, the term ``wave function'' always refers to the holomorphic part,
i.e., $\psi(z)$.

All normalizable holomorphic wave functions form a Bergman space.  The inner
product in the space is defined as
\begin{equation}
  \label{eq:inner}
  \left\langle \psi_{1} | \psi_{2} \right\rangle = \int\dd\mu_{B}
  (\bm z) \psi_{1}^{\ast}(z)\psi_{2}(z),
\end{equation}
where the integral measure is given by
\begin{equation}
  \dd{\mu_B\qty(\boldsymbol{z})}\equiv\frac{\dd[2]{\boldsymbol{z}}}{2\pi
    l_B^2}\exp(-\abs{\boldsymbol{z}}^2/2l_B^2).
\end{equation}
The measure has a Gaussian weight, which compensates for the Gaussian factors
omitted in the holomorphic wave functions.

\subsection{Reproducing kernel}
One can define the reproducing kernel
\begin{equation}
  K_{B}(z,\bar{\xi})=e^{z\bar{\xi}/2l_{B}^{2}},\label{eq:K0}
\end{equation}
which is essentially the coordinate representation of the identity operator as
well as the projection operator of the Bergman space.  The following identities
hold
\begin{align}
  \int\mathrm{d}\mu_{B}(\bm{\xi})K_{B}(z,\bar{\xi})f(\xi) & =f(z), \label{eq:project0}\\
  \int\mathrm{d}\mu_{B}(\bm{\xi})K_{B}(z,\bar{\xi})\bar{\xi}^{k}f(\xi) &
                                                                         =(2l_{B}^{2}\partial_{z})^{k}f(z). \label{project}
\end{align}
The second identity defines the projection of $\bar z$ onto the LLL, underlying
the well known rule $\bar z \rightarrow 2l_{B}^{2}\partial_{z}$~\cite{jain1997}.

\subsection{Dipole picture}
In the dipole picture, a CF is interpreted as a composite particle consisting of
an electron and a vortex.  While the electron is confined in the physical Landau
level, the vortex is assumed to reside in a separate, fictitious Landau level
generated by a Chern-Simons magnetic field oriented in the opposite direction of
the physical magnetic field.  The magnetic length of the fictitious Landau
level, $l_{b}$, is related to $l_{B}$ by
\begin{equation}
  \frac{1}{l_b^2} = \frac{2\tilde\nu}{l_B^2}.
\end{equation}
Further details of the dipole picture can be found in
Ref.~\onlinecite{shi2023quantum}.

We can also define a Bergman space as well as its reproducing kernel for the
fictitious Landau level:
\begin{align}
  \label{eq:3}
  \dd{\mu_b\qty(\boldsymbol{\eta})}&\equiv\frac{\dd[2]{\boldsymbol{\eta}}}{2\pi
                                     l_b^2}\exp(-\abs{\boldsymbol{\eta}}^2/2l_b^2), \\
  K_{b}(\bar\eta,\zeta)&=e^{\bar{\eta}\zeta/2l_{b}^{2}}. \label{eq:Kb}
\end{align}

The state of a composite fermion is thus described by a bivariate wave function
$\psi(z,\bar\eta)$, which depends on the complex electron coordinate $z$ and the
complex-conjugate vortex coordinate $\bar\eta\equiv
\eta_{x}-\mathrm{i}\eta_{y}$.

\section{From the CF binding energy to the Haffnian wave
  function \label{sec:from-cf-dispersion}}

In this section, we supplement the details of the derivation from the CF binding
energy to the physical wave function of electrons.

\subsection{Binding energy}
The CF binding energy $\epsilon_{b}(r)$, determined in the main text, is
introduced in Sec.~VB of Ref.~\onlinecite{shi2023quantum}.  It is defined such
that the energy functional of a CF can be written as
\begin{equation}
  E=\int\dd{\mu_B\qty(\boldsymbol{z})}\dd{\mu_b\qty(\boldsymbol{\eta})}
  \psi^*\qty(z,\bar\eta) \int \dd{\mu_b\qty(\boldsymbol{\eta}')} e^{\bar\eta\eta'/2l_b^2}
  \epsilon_b\qty(\boldsymbol{z};\bar\eta,\eta')\psi\qty(z,\bar\eta'),
  \label{eq_Lagrangian}
\end{equation}
where $\epsilon_b\qty(\boldsymbol{z};\bar\eta,\eta')$ is obtained from
$\epsilon_{b}(r)$ through analytic continuation by interpreting $r^{2}\equiv |\bm
z -\bm \eta |^{2}$ as $(\bar z - \bar\eta)(z-\eta^{\prime})$.

\subsection{Wave equation}
The CF wave equation can be derived by applying the variational principle of
quantum mechanics.  By minimizing the energy functional Eq.~\eqref{eq_Lagrangian} with respect to
$\psi^{\ast}(z,\bar \eta)$, subject to the normalization constraint $\int
\dd\mu_B(\bm z)\dd\mu_b(\bm\eta)|\psi(z,\bar\eta)|^{2}=1$, we obtain
\begin{equation}
  \epsilon \psi(z,\bar\eta) = \hat{H} \psi(z,\bar\eta),
  \label{eq:weq}
\end{equation}
where the Hamiltonian is given by
\begin{align}
  \left[\hat H\psi\right]\qty(z,\bar\eta) &=
                                            \hat{P}^{\prime}\epsilon_{b}(\bm
                                            z;\bar\eta,\eta^{\prime}) \psi(z,\bar\eta)\\
                                          &\equiv \int\dd{\mu_B\qty(\boldsymbol{z}')}\dd{\mu_b\qty(\boldsymbol{\eta}')}
                                            \exp(\frac{z\bar z'}{2l_B^2})\exp(\frac{\bar\eta\eta'}{2l_b^2})
                                            \epsilon_b\qty(\boldsymbol{z'};\bar\eta,\eta')\psi\qty(z',\bar\eta'), \label{eq:projectH}
\end{align}
where $\hat{P}^{\prime}$ denotes the projection onto the physical and fictitious
Landau levels.  It maps $\bar z$ and $\eta^{\prime}$ in $\epsilon_{b}(\bm z,
\bar\eta,\eta^{\prime})$ to the operators $\hat{\bar z}\equiv2l_B^2\partial_z$
and $\hat\eta\equiv2l_b^2\partial_{\bar \eta}$, respectively.

The ordering of operators after the mapping is dictated by the definition of the
projection explicitly given in Eq.~\eqref{eq:projectH}.  We see that the projections for $\bm z$
and $\bm\eta$ take different forms, corresponding to the normal ordering and the
anti-normal ordering defined in Appendix B of Ref.~\onlinecite{shi2023quantum},
respectively.  Consequently, after the projection, $\hat{\bar z} - \bar\eta$
should always be placed to the left of $z-\hat\eta$.

The rules for promoting the binding energy $\epsilon_{b}(r)$ to the Hamiltonian
$\hat H$ are summarized as follows:
\begin{itemize}
\item Express $\epsilon_{b}(r)$ as a function of $r^{2}$;
\item Replace $r^{2}$ with the operator $(2l_{B}^{2}\partial_{z} -
  \bar\eta)(z-2l_{b}^{2}\partial_{\bar\eta})$;
\item Rearrange operators so that all $(2l_{B}^{2}\partial_{z} - \bar\eta)$'s are to
  the left of $(z-2l_{b}^{2}\partial_{\bar\eta})$'s.
\end{itemize}

\subsection{Ladder operators and $\Lambda$ levels}
Ladder operators for CF $\Lambda$ levels, analogues to the ladder operators in
ordinary Landau levels, can be defined as
\begin{align}
  \hat a=&\frac{1}{\sqrt\gamma l_B}\qty(z-2l_{b}^{2}\partial_{\bar\eta}),\\
  \hat a^\dagger=&\frac{1}{\sqrt\gamma
                   l_B}\qty(2l_{B}^{2}\partial_{z} -\bar\eta),
\end{align}
for a filling factor $\tilde\nu<1/2$.  It is straightforward to verify the
commutation relation $[\hat a,\hat a^\dagger]=1$.  Additionally, one can define
operators $\hat{b}$ and $\hat{b}^{\dagger}$ to commute with $\hat{a}$ and
$\hat{a}^{\dagger}$ and relate different states within a $\Lambda$-level.  For
further details, see Appendix A of Ref.~\onlinecite{shi2023quantum}.

\begin{comment}
  \begin{align}
    \hat a=&\frac{1}{\sqrt\gamma l_B}\qty(z-\hat\eta),\\
    \hat a^\dagger=&\frac{1}{\sqrt\gamma l_B}\qty(\hat{\bar z}-\bar\eta),\\
    \hat b=&\frac{1}{\sqrt{\gamma}l_B}\qty(\frac{\hat{\bar z}}{\sqrt{2\tilde\nu}}-\sqrt{2\tilde\nu}\bar\eta),\\
    \hat b^\dagger=&\frac{1}{\sqrt{\gamma}l_B}\qty(\frac{z}{\sqrt{2\tilde\nu}}-\sqrt{2\tilde\nu}\hat\eta),
  \end{align}
\end{comment}

Hamiltonian $\hat{H}$ can then be expressed in terms of $\hat{a}$ and
$\hat{a}^{\dagger}$ by mapping $r^{2}$ to $\gamma
l_{B}^{2}\hat{a}^{\dagger}\hat{a}$ in the binding energy function
$\epsilon_{b}(r)$, and arranging $\hat{a}$ and $\hat{a}^{\dagger}$ operators
according to normal ordering.  It yields Eq.~(2) of the main text.

It is straightforward to verify that the Hamiltonian $\hat{H}$ commutes with the
number operator $\hat n =\hat a^\dagger\hat a$.  Consequently, $\Lambda$-levels
can be defined as the eigen-states of the number operator and labeled by its
eigen-values.

\subsection{Energies of $\Lambda$ levels}
It is straightforward to determine the energies of $\Lambda$ levels by
expressing $\epsilon_b\qty(r)$ as a polynomial in $r^2$.  Following the rules
outlined above, we make the substitution
\begin{equation}
  r^{2k}\to\qty(\gamma l_B^2)^k\hat a^{\dagger k}\hat a^k=\qty(\gamma
  l_B^2)^k\hat n\qty(\hat n-1)\cdots\qty(\hat n-k+1) \label{eq:quantization}
\end{equation}
to promote the polynomial to a CF Hamiltonian.  It results in a Hamiltonian
that is a function of $\hat n$~\cite{10.21468/SciPostPhys.10.1.007}.  The eigen-energy of a $\Lambda$-level can
then be obtained by substituting $\hat n$ in the Hamiltonian with the
$\Lambda$-level index $n$.

\subsection{Single-particle CF wave functions}
The wave functions of the lowest $\Lambda$ level, which has the lowest
eigen-value $n=0$ for the number operator (rather than the energy), can be
determined by solving the equation $\hat a\psi_0\qty(z,\bar\eta)=0$.  The
solution is given by
\begin{equation}
  \psi_0\qty(z,\bar\eta)=f\qty(z)\exp(\frac{z\bar\eta}{2l_b^2}),\label{eq:psi0}
\end{equation}
where $f\qty(z)$ is an arbitrary holomorphic function of $z$.  It is convenient
to choose $f_{m}(z)\propto z^{m}$, $m \in \mathbb Z^{+}$, to form a complete
basis set for the lowest $\Lambda$-level.

Wave functions for higher $\Lambda$ levels with index $n>0$ can be obtained by
applying $\hat a^\dagger$ to $\psi_0\qty(z,\bar\eta)$, as shown in the main
text.  In particular, the wave function of the second $\Lambda$-level ($n=1$)
can be written as
\begin{equation}
  \psi_n\qty(z,\bar\eta)=\hat a^{\dagger}\psi_0\qty(z,\bar\eta)\propto\exp(\frac{z\bar\eta}{2l_b^2})
  \qty(\partial_z-\frac{\bar\eta}{l^{2}}) f\qty(z), \label{eq:psi1}
\end{equation}
where $l=l_{B}/\sqrt{\tilde\nu\gamma}$ denotes the effective magnetic length of
CFs.

\subsection{Many body CF wave function}
Many body CF wave functions $\Psi_{\text{CF}}\qty(\qty{z_i,\bar\eta_i})$ are
constructed by forming Slater determinants from single-particle CF wave
functions.  For a fully-occupied lowest ($n=0$) $\Lambda$-level, using the
complete basis set defined above and Eq.~\eqref{eq:psi0}, we have:
\begin{equation}
  \Psi^0_\text{CF}\qty(\qty{z_i,\bar\eta_i})=\prod_{i<j}\qty(z_i-z_j)\prod_{i}\exp(\frac{z_i\bar\eta_i}{2l_b^2}).
\end{equation}

For the many-body CF wave function at $\nu=7/3$, where the second ($n=1$)
$\Lambda$ level is fully occupied, we apply Eq.~\eqref{eq:psi1}, and obtain:
\begin{eqnarray}
  \Psi_\text{CF}\qty(\qty{z_i,\bar\eta_i})
  &\propto&\prod_{i}\qty(2l_B^2\partial_{z_i}-\bar\eta_i)\Psi^0_\text{CF}\qty(\qty{z_i,\bar\eta_i})\\
  &\propto&\prod_{i}\exp(\frac{z_i\bar\eta_i}{2l_b^2})\qty(2\partial_{z_i}-\frac{\bar\eta_i}{l^2})\prod_{i<j}\qty(z_i-z_j).
            \label{eq_PsiCF_n1}
\end{eqnarray}

\subsection{Electron wave function}
The electron wave function is obtained by overlapping $\Psi_{\text{CF}}$ with
the $1/2$ Laughlin state of vortices~\cite{shi2023quantum}.  We have:
\begin{eqnarray}
  \Psi\qty(\qty{z_i})&=&\int\prod_{i}\dd{\mu_{b}\qty(\boldsymbol{\eta}_i)}\Psi_\text{CF}\qty(\qty{z_i,\bar\eta_i})\prod_{i<j}\qty(\eta_i-\eta_j)^2\notag\\
                     &\propto&\int\prod_{i}\dd{\mu_{b}\qty(\boldsymbol{\eta}_i)}\exp(\frac{z_i\bar\eta_i}{2l_b^2})\qty(2\partial_{z_i}-\frac{\bar\eta_i}{l^2})
                               \prod_{i<j}\qty(z_i-z_j)\qty(\eta_i-\eta_j)^2.
\end{eqnarray}
To complete the integral, we note that the exponential factor is the complex
conjugate of the reproducing kernel of the $\bm\eta$-Bergman space Eq.~\eqref{eq:Kb}.  The
integral maps $\eta_{i}$ to $z_{i}$, and $\bar\eta_{i}$ to
$2l_{b}^{2}\partial_{z_{i}}$, as indicated by Eqs.~(\ref{eq:project0}, \ref{project}).  For $\nu=7/3$,
we have $l_{b}^{2}/l^{2}=1/2$.  The electron wave function can then be written
as:
\begin{equation} \Psi\qty(\qty{z_i})\propto \lim\limits_{\{\eta_{i}\to
    z_{i}\}}\prod_{i}\qty(2\partial_{z_i}-\partial_{\eta_i})\prod_{i<j}\qty(z_i-z_j)\qty(\eta_i-\eta_j)^2.
  \label{eq_Psi}
\end{equation}

\subsection{Haffnian wave function}
To obtain an explicit form of the electron wave function, we first cast Eq. \eqref{eq_Psi}
into the form
\begin{equation}
  \Psi\qty(\qty{z_i})\propto\prod_{i<j}\qty(z_i-z_j)^3\lim\limits_{\eta\to z}\prod_{i}\qty(\hat{\mathcal D}_i+\mathcal A_i),
  \label{eq_O_prod}
\end{equation}
where we define $\hat{\mathcal D}_i\equiv2\partial_{z_i}-\partial_{\eta_i}$ and
$\mathcal A_i\equiv2\sum_{j\neq i}\qty[1/\qty(z_i-z_j)-1/\qty(\eta_i-\eta_j)]$.

We then expand the product.  Key observations are:
\begin{itemize}
\item $\lim\limits_{\eta\to z}\mathcal A_i=0$;
\item $\hat{\mathcal D}_i\mathcal A_j=4/\qty(z_i-z_j)^2+2/\qty(\eta_i-\eta_j)^2$ for
  $i\ne j$;
\item $\hat{\mathcal D}_i\hat{\mathcal D}_j\mathcal A_k=0$ for $i\ne j \ne k$.
\end{itemize}
Consequently, non-zero combinations from the expansion must pair
$\{\hat{\mathcal{D}}_{i}\}$ and $\{\mathcal{A}_{i}\} $ one by one, giving rise
to terms like $(\hat{\mathcal D}_1\mathcal A_2)(\hat{\mathcal D}_3\mathcal A_4)
\dots$ and its permutations over the indices.  Therefore, we have:
\begin{equation}
  \Psi(\{z_{i}\})	\propto \lim\limits_{\eta\to z}\sum_{\sigma\in P(N)}\prod_{i}(\hat{\mathcal D}_{\sigma\qty(2i-1)}\hat A_{\sigma\qty(2i)})\propto
  \sum_{\sigma\in P(N)}\prod_{i}\frac{1}{\qty(z_{\sigma\qty(2i-1)}-z_{\sigma\qty(2i)})^2},
  \label{eq_Hf}
\end{equation}
where the summation of $\sigma$ is over $P(N)$, the set of all permutations of $\qty{1,2,\cdots
  N}$.  The final form is recognized as the Haffnian of a matrix $\mathcal M$ with the
non-diagonal matrix elements ${\mathcal M}_{ij}=1/(z_{i}-z_{j})^{2}$, which is denoted as
$\mathrm{Hf}[1/(z_i-z_j)^2]$ in the main text.

\section{$\Lambda$-levels for $\nu=12/5$}

\begin{wrapfigure}{r}{0.40\linewidth}
  \vskip -2em
  \includegraphics[width=\linewidth]{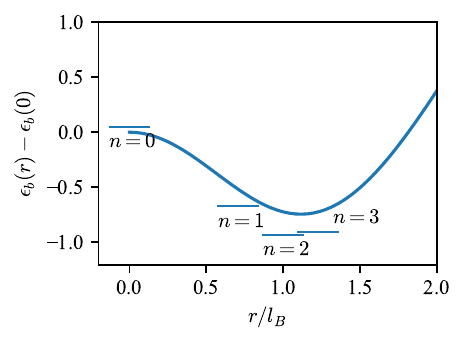}
  \caption{\label{fig:e125}Energies of $\Lambda$-levels for CF$^{2}$ in $\nu=12/5$.}
\end{wrapfigure}
According to the CF theory, the filling fraction $\nu=12/5$ ($\tilde\nu=2/5$)
corresponds to a state of CF$^{2}$ with two fully occupied $\Lambda$-levels.
Applying Eq.~\eqref{eq:quantization} with $\gamma=1/2$, we can determine the energies of
$\Lambda$-levels for this case.  The result is shown in Fig.~\ref{fig:e125},
where the two lowest-energy $\Lambda$-levels have the indices $n=2$ and $n=3$.
This differs from $\nu=2/5$, where CF$^{2}$ occupies levels $n=0$ and $n=1$.
Consequently, the ground-state wave function for $\nu=12/5$ should differ from
that for $\nu=2/5$.

Remarkably, the CF theory can predict correctly the distinction between
$\nu=12/5$ and $\nu=11/5$, despite their proximity.  Our analysis suggests that
$\nu=12/5$ is distinct from its counterpart in the LLL, whereas $\nu=11/5$ is
not, as it is a state of CF$^{4}$ with a quasiconvex dispersion.  This aligns
with observations from previous studies.

\section{Spherical geometry}

\subsection{Landau levels}
In the spherical geometry, a magnetic monopole with strength (the total number
of magnetic quantum fluxes) $2Q\in\mathbb{Z}$ is placed at the center of a
sphere, generating a uniform magnetic field over its surface.  It creates Landau
levels on the sphere, which are the eigenstates of the total angular momentum
operators $|\hat{\bm L}|^{2}$.  A Landau level, indexed by a non-negative
integer $n\ge 0$, has the angular momentum quantum number $l= Q+n$ and
degeneracy $2l+1$~\cite{jain2007composite,PhysRevLett.51.605}.

The system can be formulated in a form analogue to that of the disk geometry
using the stereographic projection~\cite{DUNNE1992233}
\begin{equation}
  z=e^{\mathrm{i}\phi} \tan\frac{\theta}{2},
\end{equation}
which maps a point with the polar angles $\qty(\theta,\phi)$ on a sphere of unit
diameter onto a complex plane.  We introduce the operator $\hat\Pi_{Q} \equiv
|\hat{\bm L}|^{2} - Q(Q+1)$.  In terms of the coordinate on the plane, the
operator can be written as
\begin{equation}
  \label{eq:Pil}
  \hat\Pi_{Q} = - \qty(1+\abs{z}^{2})^{2}\qty(\partial_{z}-Q\frac{\bar{z}}{1+\abs{z}^{2}})\qty(\partial_{\bar{z}}+Q\frac{z}{1+\abs{z}^{2}}).
\end{equation}
Landau levels are eigenstates of $\hat\Pi_{Q}$, with eigenvalues
$l(l+1)-Q(Q+1)$.

The wave functions of the LLL take the general form
\begin{equation}
  \label{eq:wfsphr}
  \psi^{(Q)}_{0}(z) \left(1+|z|^{2}\right)^{-Q},
\end{equation}
where $\psi^{(Q)}_{0}(z)$ is a holomorphic polynomial in $z$ with a degree no
greater than $2Q$.  The factor $(1+|z|^{2})^{-Q}$ is analogous to the Gaussian
factor in the wave functions of the disk geometry.  A complete basis set for the
LLL is given by the polynomials:
\begin{equation}
  \psi^{(Q)}_{0,m}(z) \propto z^m,\ m=0,1,2,\cdots 2Q. \label{eq:psi0m}
\end{equation}

The wave functions of higher Landau levels can be constructed from those of the
LLL.  For a system with monopole strength $2Q$, the Landau level with the index
$n$ has the degeneracy $2(Q+n)+1$, identical to the degeneracy of the LLL for a
monopole strength $2(Q+n)$.  This allows for an one-to-one mapping between their
states.  The wave function in the $n$-th Landau level for monopole strength $2Q$
takes the form
\begin{equation}
  \label{eq:6}
  \psi_{n}^{(Q)} (\bm z) \left(1+|z|^{2}\right)^{-(Q+n)}.
\end{equation}
It can be related to the LLL wave function $\psi_{0}^{(Q+n)}(z)$ for monopole
strength $2(Q+n)$ by
\begin{equation}
  \label{eq:mapping}
  \psi_{n}^{(Q)}(\bm z) \propto\left( \prod_{i=1}^{n}\hat
    A_{2Q+n+i} \right) \psi_{0}^{(Q+n)}(z),
\end{equation}
where
\begin{equation}
  \label{eq:5}
  \hat{A}_{q} \equiv (1+|z|^{2})\partial_{z} - q \bar z,\,\, q \in \mathbb Z^{+},
\end{equation}
acts as a raising operator for the spherical geometry.  The following identity
holds:
\begin{equation}
  \label{eq:1}
  \hat\Pi_{q}\hat A_{q+1} = \hat A_{q+1} \hat\Pi_{q+1} + 2(q+1).
\end{equation}
Using the identity, it is straightforward to show that the wave function Eq.~\eqref{eq:wfsphr}
constructed using Eq.~\eqref{eq:mapping} is indeed an eigenstate of $\hat\Pi_{Q}$ with the
expected eigenvalue for the $n$-th $\Lambda$ level.

\subsection{Bergman space}
We can define the Bergman space for the LLL on a sphere.  It has the integral
measure:
\begin{equation}
  \dd{\mu_{Q}\qty(\boldsymbol{z})}=\frac{\qty(2Q+1)\dd[2]{\boldsymbol{z}}}{\pi}\qty(\frac{1}{1+\abs{z}^{2}})^{2Q+2},
\end{equation}
and the reproducing kernel
\begin{equation}
  K_Q\qty(z,\bar \xi)=\qty(1+z\bar \xi)^{2Q}. \label{eq:Kl}
\end{equation}

Projection identities on a sphere analogue to Eq.~\eqref{project} read
\begin{align}
  \int\mathrm{d}\mu_{Q}(\bm{\xi})K_{Q}(z,\bar{\xi})\frac{\bar{\xi}^{m}}{\left(1+\left|\xi\right|^{2}\right)^{m}}\psi(\xi) & =\frac{(2Q+1)!}{(2Q+1+m)!}\partial_{z}^{m}\psi(z),\label{eq:prjspr1}\\
  \int\mathrm{d}\mu_{Q}(\bm{\xi})K_{Q}(z,\bar{\xi})\frac{\bar{\xi}^{m}}{\left(1+z\bar{\xi}\right){}^{m}}\psi(\xi) & =\frac{(2Q-m)!}{(2Q)!}\partial_{z}^{m}\psi(z).\label{eq:prjspr2}
\end{align}

\begin{comment}
  \begin{equation}
    \int\dd{\mu_{l}\qty(\boldsymbol{\xi})}K_{l}(z,\bar\xi)\qty(\frac{\bar \xi}{1+z\bar \xi})^kf\qty(\xi)
    =\frac{(2l-k)!}{(2l)!}\partial_{z}^k f\qty(z).
    \label{eq_proj_sphere}
  \end{equation}
\end{comment}
\begin{comment}
  \begin{figure}[t]
    \centering
    \includegraphics[width=0.6\linewidth]{./fig/stereographic}
    \caption{Sketch of stereographic projection for electron LLs. }
  \end{figure}
\end{comment}

\subsection{$\Lambda$ levels on a sphere}
For the dipole model of spherical geometry, we assume a monopole field strength
$2Q$ for electrons and a negative monopole strength $-2q=-2(N-1)$ for vortices.

In the spherical geometry, just like the disk geometry, $\Lambda$-levels can be
defined as the eigen-states of the $\hat r^{2}$ operator, where $r^{2}$ is
interpreted as the squared chord distance between an electron and a vortex on
the surface of the sphere.  The squared chord distance can be expressed in terms
of the stereographic coordinates of the electron and vortex:
\begin{equation}
  r^2=\frac{\abs{z-\eta}^2}{\qty(1+\abs{z}^2)\qty(1+\abs{\eta}^2)},
\end{equation}
The $\hat r^{2}$ operator is defined through a projection to the CF Bergman
space:
\begin{equation}
  \hat{r}^{2}\psi\qty(z,\bar{\eta})\equiv\int\dd{\mu_{Q}\qty(\bm{\xi})}\dd{\mu_{q}\qty(\boldsymbol{\eta}')}K_{Q}(z,\bar{\xi})K_{q}(\bar{\eta},\eta^{\prime})\frac{(\bar{\xi}-\bar{\eta})(\xi-\eta^{\prime})}{(1+|\xi|^{2})(1+\bar{\eta}\eta^{\prime})}\psi\qty(\xi,\bar{\eta}^{\prime}).
\end{equation}
By applying the projection identities Eqs.~(\ref{eq:prjspr1}, \ref{eq:prjspr2}), we obtain the explicit
form of the operator:
\begin{equation}
  \hat{r}^{2}=-\frac{\qty(1+z\bar{\eta})^{2}}{4(Q+1)q}\qty(\partial_{z}-2Q\frac{\bar{\eta}}{1+z\bar{\eta}})\qty(\partial_{\bar{\eta}}-2q\frac{z}{1+z\bar{\eta}}).\label{eq_HCF_sphere}
\end{equation}

The $\hat r^{2}$ operator can be related to the $\Pi_{\mathcal Q}$ operator for
ordinary Landau levels with monopole strength $2\mathcal Q\equiv2(Q-q)$ via an
orthogonal transformation combined with analytic continuation:
\begin{equation}
  \hat\Pi_{\mathcal Q}  \propto\eval{(1+z\bar{\eta})^{-Q-q}\hat{r}^{2}\qty(1+z\bar{\eta})^{Q+q}}_{\bar{\eta}\to\bar{z}}
\end{equation}

Thus, in the spherical geometry, similar to the disk geometry (see Appendix A of
Ref.~\cite{shi2023quantum}), $\Lambda$-level wave functions can be inferred from their
Landau-level counterparts using the relation:
\begin{equation}
  \psi_{nm}(z,\bar\eta) \propto \left. \psi^{(\mathcal Q)}_{nm}(z,\bar z)
  \right|_{\bar z \rightarrow \bar\eta} \left(1+z\bar\eta\right)^{2q-n},  \label{eq:Lwf}
\end{equation}
where $\psi_{nm}(z,\bar \eta)$ denotes a wave function of the $n$-th
$\Lambda$-level, and $\psi^{(\mathcal Q)}_{nm}(z, \bar z)$ a wave function of
the $n$-th Landau level.

\subsection{Many body CF wave function}
Using Eqs.~(\ref{eq:psi0m}, \ref{eq:mapping}, \ref{eq:Lwf}), we can construct a complete basis set of the second
$\Lambda$-level as follows:
\begin{align}
  \psi_{1,m}(z,\bar{\eta}) & \propto\left(1+z\bar{\eta}\right)^{2q-1}\left[(1+z\bar{\eta})\partial_{z}-2(\mathcal{Q}+1)\bar{\eta}\right]z^{m}\\
                           & =K_{q}(z,\bar{\eta})\left(\partial_{z}-\frac{2(\mathcal{Q}+1)\bar{\eta}}{1+z\bar{\eta}}\right)z^{m},
\end{align}
where we $K_{q}(z,\bar\eta)$ is the reproducing kernel defined in Eq.~\eqref{eq:Kl}.

The many body CF wave function $\Psi_{\text{CF}}\qty(\qty{z_i,\bar\eta_i})$ for
a fully occupied second ($n=1$) $\Lambda$-level is then written as
\begin{equation}
  \Psi_{\text{CF}}\qty(\qty{z_{i},\bar{\eta}_{i}})\propto\prod_{i}K_{q}(z_{i},\bar{\eta}_{i})\left[\partial_{z_{i}}-\frac{2(\mathcal{Q}+1)\bar{\eta}_{i}}{1+z_{i}\bar{\eta}_{i}}\right]\prod_{i<j}\qty(z_{i}-z_{j}).
\end{equation}

\subsection{Electron wave function}
The corresponding electron wave function is given by
\begin{equation}
  \Psi\qty(\qty{z_{i}})\propto\int\prod_{i}\dd{\mu_{q}\qty(\boldsymbol{\eta}_{i})}K_{q}(z_{i},\bar{\eta}_{i})\left[\partial_{z_{i}}-\frac{2(\mathcal{Q}+1)\bar{\eta}_{i}}{1+z_{i}\bar{\eta}_{i}}\right]\prod_{i<j}\qty(z_{i}-z_{j})\qty(\eta_{i}-\eta_{j})^{2}.
\end{equation}
By applying Eq.~\eqref{eq:prjspr2}, it simplifies to
\begin{equation}
  \Psi(\{z_{i}\})\propto\lim_{\{\eta_{i}\rightarrow z_{i}\}}\prod_{i}\left[\partial_{z_{i}}-\frac{\mathcal{Q}+1}{q}\partial_{\eta_{i}}\right]\prod_{i<j}(z_{i}-z_{j})(\eta_{i}-\eta_{j})^{2}.
\end{equation}
For a fully occupied second $\Lambda$-level, we have $2\mathcal{Q}+3=N$, and
using the identity $q=\qty(N-1)$, it follows $(\mathcal{Q}+1)/q = 1/2$.  The
wave function then reduces to a form that is identical to Eq.~\eqref{eq_Psi}.
\begin{comment}
  \begin{eqnarray}
    \Psi\qty(\qty{z_i})&=&\int\prod_i\dd{\mu_{N-1}\qty(\boldsymbol{\eta}_i)}\Psi^\text{CF}\qty(\qty{z_i,\bar\eta_i})\prod_{i<j}\qty(\eta_i-\eta_j)^2\notag\\
	&\propto&\int\prod_i\dd{\mu_{N-1}\qty(\boldsymbol{\eta}_i)}\prod_{i}\qty(\frac{4R^2}{3N-3}\partial_{z_i}-\frac{\bar\eta_i}{1+z_i\bar\eta_i/4R^2})
           \qty(1+\frac{z_i\bar\eta_i}{4R^2})^{2N-2}
           \prod_{i<j}\qty(z_i-z_j)\qty(\eta_i-\eta_j)^2\notag\\
	&\propto&\int\prod_i\dd{\mu_{N-1}\qty(\boldsymbol{\eta}_i)}
           \prod_{i}\qty(1+\frac{z_i\bar\eta_i}{4R^2})^{2N-2}
           \qty(4R^2\partial_{z_i}-\frac{\qty(N-1)\bar\eta_i}{1+z_i\bar\eta_i/4R^2})
           \prod_{i<j}\qty(z_i-z_j)\qty(\eta_i-\eta_j)^2\notag\\
	&=&\prod_{i<j}\qty(z_i-z_j)\int\prod_i\dd{\mu_{N-1}\qty(\boldsymbol{\eta}_i)}\prod_{i}
           \qty(4R^2\overleftarrow{\partial_{z_i}}-\frac{\qty(N-1)\bar\eta_i}{1+z_i\bar\eta_i/4R^2})\qty(1+\frac{z_i\bar\eta_i}{4R^2})^{2N-2}
           \prod_{i<j}\qty(\eta_i-\eta_j)^2\notag\\
	&=&\prod_{i<j}\qty(z_i-z_j)\int\prod_i\dd{\mu_{N-1}\qty(\boldsymbol{\eta}_i)}\prod_{i}
           \qty(4R^2\overleftarrow{\partial_{z_i}}-2R^2\partial_{z_i})\qty(1+\frac{z_i\bar\eta_i}{4R^2})^{2N-2}
           \prod_{i<j}\qty(\eta_i-\eta_j)^2\notag\\
	&\propto&\lim\limits_{\eta\to z}\prod_{i}\qty(2\partial_{z_i}-\partial_{\eta_i})\prod_{i<j}\qty(z_i-z_j)\qty(\eta_i-\eta_j)^2.
  \end{eqnarray}
\end{comment}

\section{Exact diagonalizations}
\subsection{Disk geometry}
We simulate a system of $N$ interacting electrons on a disk of radius
$R_{c}\equiv\sqrt{2N/\tilde\nu}$ (with $l_{B}=1$).  The disk contains a
neutralizing positive charge background with a uniform charge density within
$R_{c}$.  The presence of the neutralizing charge background induces a
single-body potential, $\Phi(r)$, experienced by electrons.

Exact diagonalizations are performed in a finite Hilbert space spanned by the
basis set $\phi_{m}(z)=z^{m}/\sqrt{m!}$, with $m<{\tilde\nu}^{-1}N+D$ and $D\in
\mathbb Z^{+}$.  This allows electrons to extend beyond the disk radius by $D$
additional orbits.  For the results presented in the main text, we set $D=5$.

The interaction matrix elements can be determined analytically.  For the LLL,
the explicit form for the Coulomb interacting matrix elements,
$V_{m_1m_2m_3m_4}$, is provided in Ref.~\onlinecite{tsiper2002analytic}.
\begin{comment}
  \begin{equation}
    V_{m_1m_2m_3m_4}^{(n)}\equiv\int\dd{\mu_B\qty(\boldsymbol{z})}\dd{\mu_B\qty(\boldsymbol{z}')}
    \psi_{nm_1}^*\qty(\boldsymbol{z})\psi_{nm_3}\qty(\boldsymbol{z})v_\mathrm{c}\qty(\boldsymbol{z}-\boldsymbol{z})
    \psi_{nm_2}^*\qty(\boldsymbol{z}')\psi_{nm_4}\qty(\boldsymbol{z}')\label{eq_Vmel_def},
  \end{equation}
\end{comment}
For the 1LL, the matrix elements $\tilde V_{m_1m_2m_3m_4}$ of the
effective interaction can be expressed in terms of $V_{m_1m_2m_3m_4}$ as follows:
\begin{eqnarray}
  &&\tilde V_{m_1m_2m_3m_4}=\notag\\
  &&\sqrt{\qty(m_1+1)\qty(m_3+1)}\left(\sqrt{\qty(m_2+1)\qty(m_4+1)}V_{m_1+1,m_2+1,m_3+1,m_4+1}
     -\qty(m_2+m_4)V_{m_1+1,m_2,m_3+1,m_4}\right.\notag\\
  &&\qquad\left.+\sqrt{m_2m_4}V_{m_1+1,m_2-1,m_3+1,m_4-1}\right)\notag\\
  &&-\qty(m_1+m_3)\qty(\sqrt{\qty(m_2+1)\qty(m_4+1)}V_{m_1,m_2+1,m_3,m_4+1}
     -\qty(m_2+m_4)V_{m_1m_2m_3m_4}+\sqrt{m_2m_4}V_{m_1,m_2-1,m_3,m_4-1})\notag\\
  &&+\sqrt{m_1m_3}\left(\sqrt{\qty(m_2+1)\qty(m_4+1)}V_{m_1-1,m_2+1,m_3-1,m_4+1}
     -\qty(m_2+m_4)V_{m_1-1,m_2,m_3-1,m_4}\right.\notag\\
  &&\qquad\left.+\sqrt{m_2m_4}V_{m_1-1,m_2-1,m_3-1,m_4-1}\right).
\end{eqnarray}
The relation can be derived from the connection between the single-particle
electron wave function $\psi_{1m}\qty(\boldsymbol{z})$ in the 1LL and its
counterparts in the LLL $\psi_{0m}\qty(z)\equiv \phi_{m}(z)$:
$\psi_{1m}\qty(\boldsymbol{z})=\bar{z}\psi_{0m}\qty(z)/\sqrt{2}-\sqrt{m}\psi_{0,m-1}\qty(z)$.
\begin{comment}
  We apply the identity:
  \begin{eqnarray}
    \psi_{1m}^*\qty(\boldsymbol{z})\psi_{1m'}\qty(\boldsymbol{z})&=&
           \sqrt{\qty(m+1)\qty(m'+1)}\psi_{0,m+1}^*\qty(z)\psi_{0,m'+1}\qty(z)\notag\\
	&&-\qty(m+m')\psi_{0m}^*\qty(z)\psi_{0m'}\qty(z)
           +\sqrt{mm'}\psi_{0,m-1}^*\qty(z)\psi_{0,m'-1}\qty(z)
  \end{eqnarray}
  to Eq. \eqref{eq_Vmel_def}
\end{comment}

Hamiltonian matrices are constructed using the interaction matrix elements and
the single-particle potential $\Phi(r)$, which is diagonal in the basis set
$\{\phi_{m}\}$.  The resulting matrices are then diagonalized using the Lanczos
algorithm.

It is necessary to test our calculations under different simulation setups, as
the disk geometry introduces boundary effects that may influence the results.
Two factors can be identified: (i) the charge distribution of the neutralizing
background near the boundary, and (ii) the value of $D$ which determines the
number of basis states used in ED.  For (i), we smear the abrupt
change in the neutralizing charge density at $R_{c}$ by assuming that the
background has the charge distribution:
\begin{equation}
  \rho_{b}\qty(r)=\rho_{0}\frac{\Gamma\qty(\frac{R_{c}^{2}}{2\lambda^{2}},\frac{r^{2}}{2\lambda^{2}})}{\Gamma\qty(\frac{R_{c}^{2}}{2\lambda^{2}})},
\end{equation}
where $\Gamma(x,y)$ is the upper incomplete gamma function, and $\Gamma(x)\equiv
\Gamma(x,0)$ is the complete gamma function.  This results in a smooth decrease
in density near the boundary over a length scale $\sim\lambda$.  For (ii), we
perform calculations for various values of $D$.

\begin{figure}[h]
  \centering
  \includegraphics[width=0.8\linewidth]{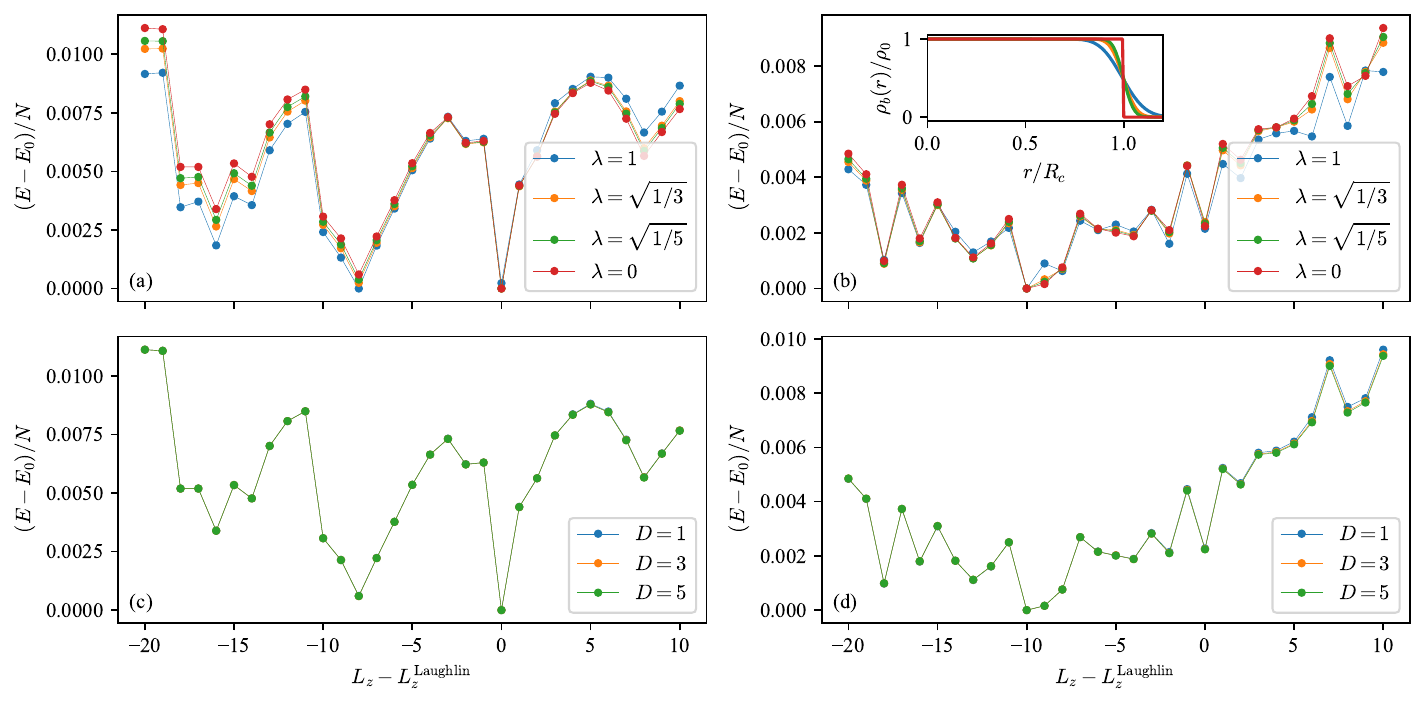}
  \caption{ED ground state energies for different simulation parameters with
    $N=10$.  (a) and (b) show results for varying values of $\lambda$, while (c)
    and (d) present results for different values of $D$.  Results for the LLL
    are displayed in the left [(a) and (c)], and for the 1LL in the right [(b)
    and (d)].}
  \label{fig_supp_disk_test}
\end{figure}

ED results for various values of $\lambda$ and $D$ are shown in Fig.~\ref{fig_supp_disk_test}.  It
is evident that these parameters have only minor effects on the results.  Our
main conclusion that the ground states for $\nu=1/3$ and $\nu=7/3$ have
different expectation values of $L_{z}$ remains unaffected by the choice of
simulation parameters.

\subsection{Spherical geometry}
Simulating the system on a sphere removes the need for a
neutralizing charge background and eliminates boundary effects, making it
preferable to disk simulations.  However, electron states on a sphere depends
on the topological shift $S$, requiring us to test various $(N,S)$
pairs.  The results are presented in the main text.

Hamiltonian matrices are constructed using the Coulomb interaction matrix
elements, for both the LLL and 1LL, as given in Ref.~\onlinecite{jain2007composite}.  Resulting
Hamiltonian matrices are then diagonalized using the Lanczos algorithm.

\section{CF$^2$ in a finite-width quantum well}

\begin{figure}[htb]
  \centering
  \includegraphics[width=0.7\linewidth]{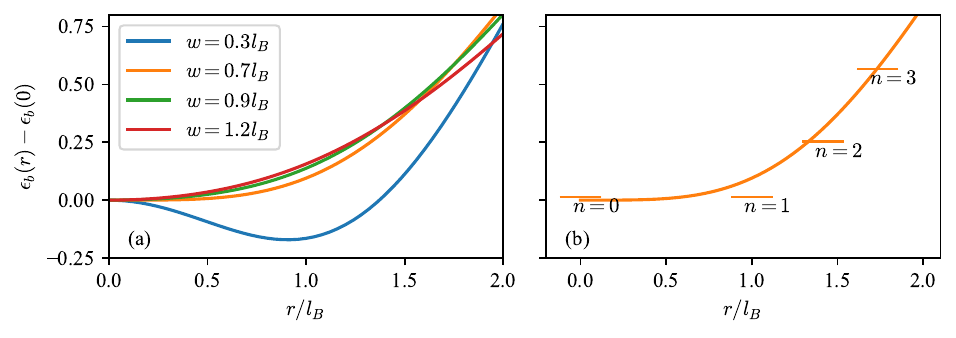}
  \caption{(a) Dispersions of CF$^2$ in the 1LL with different values of
    $w$. (b) Dispersion and energies of $\Lambda$ levels with $w=0.7l_B$.  The
    energies are in units of $\tilde{\nu} e^2/16\pi^{2} \epsilon l_B$.}
  \label{fig_supp_finite_width}
\end{figure}

In Fig.~\ref{fig_supp_finite_width}, we present the dispersions of CF$^2$ in a finite-width quantum well for
different values of the width parameter $w$ (see the main text).  As $w$
increases, the dispersion of CFs evolves from non-quasiconvex to quasiconvex.
The transition occurs at $w \approx 0.7 l_B$,  as shown in the right panel of
Fig. \ref{fig_supp_finite_width}, where the energies of the first and second $\Lambda$ levels becomes degenerate.

\section{Self-consistent determination of the CF dispersion}

In the main text, we determine the dispersions of CF$^{2}$ for both the LLL and
1LL by assuming that electrons are in the $1/3$ Laughlin state.  While this is
self-consistent for the LLL, it is not for the 1LL, as its CF dispersion
suggests that $\nu=7/3$ corresponds to a fermionic Haffnian state.  Ideally, to
achieve full self-consistency, the dispersion should also be determined using
the Haffnian state.

Following the mean-field approach developed in Sec.~VB of
Ref.~\onlinecite{shi2023quantum}, we determine the electron wave function in
the presence of a vortex at the origin:
\begin{eqnarray}
  \Psi_0^{\mathrm v}\qty(\qty{z_i})&=&
                                       \int\prod_i\dd{\mu\qty(\boldsymbol{\eta}_i)}\eta_i^2\Psi_\text{CF}\qty(\qty{z_i,\bar\eta_i})\prod_{i<j}\qty(\eta_i-\eta_j)^2\notag\\
                                   &\propto& \lim\limits_{\{\eta_{i}\to
                                             z_{i}\}}\prod_{i}\qty(2\partial_{z_i}-\partial_{\eta_i})\prod_{i<j}\qty(z_i-z_j)\qty(\eta_i-\eta_j)^2
                                             \prod_{i} \eta_{i}^{2},
\end{eqnarray}
where $\Psi_\text{CF}\qty(\qty{z_i,\bar\eta_i})$ is given by Eq.~\eqref{eq_PsiCF_n1}.  The
electron density profile in the vicinity of the vortex can then be determined.

Following the steps outlined in Sec.~\ref{sec:from-cf-dispersion}, we obtain the explicit form of
the wave function:
\begin{equation}
  \Psi_0^{\mathrm v}\qty(\qty{z_i}) \propto \mathrm{lHf}(\mathcal M) \prod_{i}z_i^2\prod_{i<j}\qty(z_i-z_j)^3\label{eq_Psi_lHf},
\end{equation}
where $\mathcal M$ is a matrix with elements $\mathcal M_{ij}=1/(z_i-z_j)^2$ for
$i\neq j$, and $\mathcal M_{ii}=-\sqrt{2/3}/z_i$.  $\mathrm{lHf}(\mathcal M)$
denotes the loop Haffnian of the matrix~\cite{bjorklund2019fasterhafnianformulacomplex}.  The particular loop Haffnian can
also be written as a determinant: $\mathrm{lHf}({\mathcal
  M})=\mathrm{det}(\tilde{\mathcal M})$, where $\tilde{\mathcal M}$ is a matrix
with elements $\tilde{\mathcal M}_{ij}=1/(z_i-z_j)$ for $i\neq j$, and
$\tilde{\mathcal M}_{ii}=-\sqrt{2/3}/z_i$.

\begin{figure}[bt]
  \centering
  \includegraphics[width=0.666\linewidth]{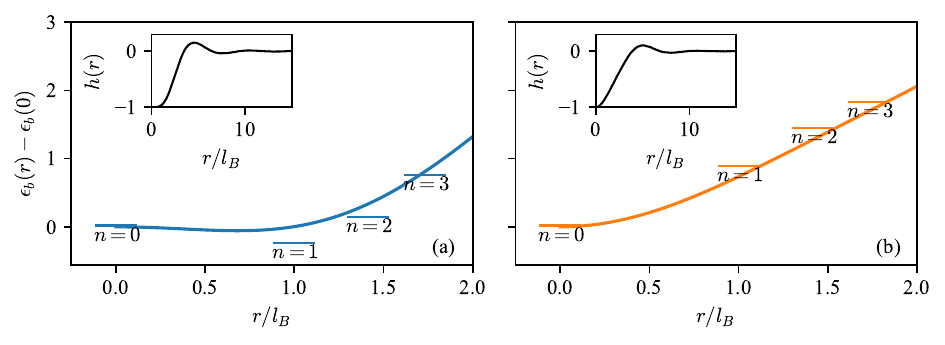}
  \caption{(a) $h(r)$, $\epsilon_b\qty(r)$ and $\Lambda$ levels of the $7/3$
    state based on Eq. \eqref{eq_Psi_naive}.  (b) Same as (a) but
    $\Psi_0^{\mathrm v}\qty(\qty{z_i})$ is given by Eq. \eqref{eq_Psi_lHf}.  The
    energies are in units of $\tilde{\nu} e^2/16\pi^{2} \epsilon l_B$.}
  \label{fig_supp_self_consistent}
\end{figure}

In Fig.~\ref{fig_supp_self_consistent}, we present the CF dispersion and $\Lambda$ level energies computed
using the loop Haffnian wave function.  The dispersion is quasiconvex and
exhibits a steep increase with $r$.  This is markedly different from the dispersion
obtained when assuming the Laughlin wave function.

We attribute the failure to the inadequacy of the mean-field approach, which
neglects the exchange symmetry between CFs.  The resulting absence of Pauli
exclusion induces the issue, which is further amplified by the pairing nature of
the Haffnian state.  To illustrate, consider the expansion of the loop Haffnian by its
diagonal elements:
\begin{equation}
  \mathrm{lHf}\qty(\mathcal M)=\sum_{s=0}^{N} \sum_{\{i_{1},i_{2}\cdots
    i_{s}\}}\frac{1}{z_{i_1}}\cdots\frac{1}{z_{i_s}}\times\mathrm{Hf}\qty(m_{\{i_{1},i_{2}\cdots
    i_{s}\}}), \label{eq:lhfwf}
\end{equation}
where $\{i_1,i_2\cdots,i_s\}$ denotes a list of distinct indices,
$m_{\{i_{1},i_{2}\cdots i_{s}\}}$ is the matrix obtained by removing the columns
and rows with indices $i_1,i_2\cdots i_s$ in $\mathcal M$.  The factor
$1/z_{i_1}\cdots 1/z_{i_s}$ indicates that there are $s$ additional electrons
which have pairing correlations with the vortex at the origin, despite the
vortex already being bound to an electron which is not included in
$\Psi^{\mathrm v}_{0}$, according to the assumption of the mean-field approach.
This is obviously an artifact that arises from the lack of Pauli-exclusion
between the vortex-bound electron and other electrons in the system.  Its impact
is exacerbated by the $1/z$ correlation, which is absent in the Laughlin wave
function. 
\begin{comment}
  \begin{eqnarray}
    \Psi_0^{\mathrm v}\qty(\qty{z_i})&=&
           \int\prod_i\dd{\mu\qty(\boldsymbol{\eta}_i)}\eta_i^2\Psi^\text{CF}\qty(\qty{z_i,\bar\eta_i})\prod_{i<j}\qty(\eta_i-\eta_j)^2\notag\\
	&\propto&\lim\limits_{\eta\to z}\prod_{i}\qty(2\partial_{z_i}-\partial_{\eta_i})\prod_{i<j}\qty(z_i-z_j)\qty(\eta_i-\eta_j)^2\prod_{i}\eta_i^2\notag\\
	&=&\prod_{i}z_i^2\prod_{i<j}\qty(z_i-z_j)^3\lim\limits_{\eta\to z}\prod_{i}\qty(\hat{\mathcal D}_i+\mathcal A_i-\frac{2}{\eta_i})\notag\\
	&\propto&\prod_{i}z_i^2\prod_{i<j}\qty(z_i-z_j)^3\mathrm{lHf}(\mathcal M_{ij})\label{eq_Psi_0v}
  \end{eqnarray}
\end{comment}
\begin{comment}
  $h\qty(r)$ is computed using the wave function $\Psi_0^\mathrm{v}\qty(\qty{z_i})=\det\qty(P_{ij})\prod_{i}z_i^2 \prod_{i<j}\qty(z_i-z_j)^3$, where $P_{ij}=1/\qty(z_i-z_j)$ for $i\neq j$ and $P_{ii}=-\sqrt{2/3}/z_i$.  The detail is discussed in supplemental material.  We find that $n=1$ $\Lambda$ state is still the lowest.  In this way we predict that $7/3$ is fermionic Hafnian state.
\end{comment}

To estimate the CF dispersion for the Haffnian state, we discard the
terms with $s\ne 0$ in the expansion Eq.~\eqref{eq:lhfwf}, as these are considered unphysical.
The resulting wave function with a vortex is:
\begin{equation}
  \Psi_{0}^{\mathrm v}\qty(\qty{z_i})=\prod_{i}z_i^2\Psi^{\text{Hf}}\qty(\qty{z_i})\label{eq_Psi_naive},
\end{equation}
where $\Psi^{\text{Hf}}\qty(\qty{z_i})$ is the fermionic Haffnian wave function
(Eq. [4] of the main text).  The corresponding CF dispersion and $\Lambda$ level
energies are shown in Fig. \ref{fig_supp_self_consistent} (a).  The dispersion determined using the
Haffnian wave-function is indeed non-quasiconvex, although the minimum is
significantly shallower compared to that determined using the Laughlin wave
function.

Properly determining the CF dispersion for the Haffnian state may require an
improved mean field approach, e.g., a Hartree-Fock theory for CFs, which fully
accounts for the exchange symmetry.

\bibliographystyle{apsrev4-2.bst} \bibliography{ref}